\input{aipcheck}
\documentclass[final]{aipproc}

\layoutstyle{8x11single}

\def\p{\mathrm{p}}

\def\b{\mathrm{b}}
\def\d{\mathrm{d}}
\def\v{\mathrm{v}}
\def\m{\mathrm{m}}
\def\mp{{\mathrm{m}^\prime}}

\def\vr{\vec{r}}
\def\vn{\vec{n}}
\def\vv{\vec{v}}
\def\ve{\vec{E}}
\def\vf{\vec{F}}
\def\va{\vec{A}}
\def\vb{\vec{B}}
\def\vp{\vec{p}}
\def\vnabla{\vec{\nabla}}

\def\cf{{\cal F}}

\begin{document}

\title{Extracting  physical observables using  macroscopic ensemble in the spex-mixer/mill simulation}


\author{Muhandis\thanks{muhandis@teori.fisika.lipi.go.id}}{
  address={Group for Theoretical and Computational Physics, Research Center for Physics, Indonesian Institute of Sciences (LIPI), Kompleks Puspiptek Serpong, Tangerang 15310, Indonesia}
}
\author{F. Nurdiana}{
  address={Department of Physics, University of Indonesia, Kampus UI Depok, Depok 16424, Indonesia}
}
\author{A.S. Wismogroho}{
  address={Group for Nanomaterial, Research Center for Physics, Indonesian Institute of Sciences (LIPI), Kompleks Puspiptek Serpong, Tangerang 15310, Indonesia}
}
\author{N.T. Rochman}{
  address={Group for Nanomaterial, Research Center for Physics, Indonesian Institute of Sciences (LIPI), Kompleks Puspiptek Serpong, Tangerang 15310, Indonesia}
}
\author{L.T. Handoko\thanks{laksana.tri.handoko@lipi.go.id, handoko@teori.fisika.lipi.go.id}}{
  address={Group for Theoretical and Computational Physics, Research Center for Physics, Indonesian Institute of Sciences (LIPI), Kompleks Puspiptek Serpong, Tangerang 15310, Indonesia}
  ,altaddress={Department of Physics, University of Indonesia, Kampus UI Depok, Depok 16424, Indonesia}
}

\begin{abstract}
A technique to simulate the spex-mixer/mill system as a macroscopic ensemble rather than a pure dynamical system is proposed. The treatment is suitable especially for comminution processes generating the nanomaterial up to  nanometers scale where the geometrical displacements are unobservable. It is argued that the method is simple and experimentally verifiable since relevant physical observables can be extracted using partition function without solving the equation of motions.
\end{abstract}

\maketitle

\section{Introduction}
\label{intro}

The comminution processes in recent years attract the attention among scientists and engineers due to the increasing demand of ultrafine powders for nanotechnology applications in many areas. The demand then requires the improvement of comminution equipments like ball mills, roller mills and so on. Unfortunately, the development of such comminution equipments always contains a lot of uncertainties due to a wide range of unknown parameters. These, in fact, lead to significant statistical errors. In order to overcome such problems, several models have been developed to quantitatively describe comminution process in various types of mills \cite{mishra92,mishra94,mishra95,mishra01,poschel}.

On the other hand, mathematical modeling and simulation may provide prior information and constraint to the unknown parameter ranges which should be useful to develop more optimized experimental strategy in comminution processes. However, in most cases of mathematical models, the physical observables like grain-size etc are extracted from a set of equation of motions (EOM). Such EOM's are considered to govern as complete as possible the dynamics of the system, from the mechanical motions to the evolution of grain-size distribution. This approach is obviously suffered from the nonlinearities of the equations under consideration, and then the requirement of high computational power to solve them numerically. This fact often discourages a quantitative and deterministic approach for the simulation of such system. These nonlinear effects like chaotic behavior of the sphere motions within the mill encourages some works modeling the system using semi-empirical approaches \cite{manai}. However most of semi-empirical models require a large number of experimental data based on prior observations \cite{davis}, or measured variables obtained from simulation results by other authors \cite{maurice1990,maurice1996}.

More empirical approach is based on the physically realistic modelization of the ball mill system \cite{delogu,wang}. This approach in general deals with three aspects : (1) evaluation of milling bodies dynamics and energetic inputs transferred to powders; (2) description of the effects of such inputs on powders breakage; (3) description of powders evolution in terms of particle size distribution \cite{concas}. In a recent work \cite{concas}, a comprehensive study on this line for the case of spex mixer / mill was  performed by deploying the 3D simulation for milling bodies motion, and the population balance method to describe the granulometric evolution. Then, both are related through the energetic inputs in the population balance formalism which is obtained from the simulation of milling bodies motion.

In this paper we propose a novel model and approach combining the deterministic approach for milling bodies motion, and the statistical approach to relate them with considerable macroscopic physical observables. This work is devoted to overcome the following problems in conventional approach :
\begin{itemize}
\item Experimentally it is almost impossible to trace the geometrical displacements of all  matters in a vial with proper time resolution to verify the models which are based on the classical EOM. This problem is getting worse as one simulates a system with matters at few nanometers scale with comparable size of time-space resolution.
\item Solving a set of EOMs numerically, and then performing a simulation with high accuracy (enough time resolution) require huge efforts on both computing capacity and running time.
\item Taking into account the external circumstances around the vial like electromagnetic field and so forth. This might be interesting when one considers a comminution process which can reach the level of few nanometers.
\end{itemize}
Therefore this work is intended to provide a tool for a nanometer system in a vial by developing direct relations between the vial internal dynamics with some external physical observables which should be easier to measure. We should remark here that the vial internal dynamics is yet described empirically using physical modelization approach.

Further, rather solving the EOM's governing the whole dynamics, we use the hamiltonian approach to model all relevant interactions, and extract the physical observables through partition function by considering the system as a canonical ensemble in finite temperature. As a consequence, instead of observing the geometrical evolution of matters in term of time in a ball mill, we can investigate the particle number distribution in term of temperature. This introduces a novel method relating the internal dynamics with the macroscopic physical parameter like temperature, rather than time and geometrical displacements which are in most cases difficult to realize. An example of numerical simulation is given for the case of ball mill with a structure similar to the  well-known spex mixer / mill.

\section{The model}

The whole system is modeled empirically using hamiltonian method. First we construct the total hamiltonian describing the dynamics in the ball mills. It is further followed by formulating the partition function and extracting the relevant thermodynamics observables.

\subsection{The dynamics}

In our model, the dynamics of each 'matter' in the system, i.e. balls and powders inside the vial, is described by a hamiltonian $H_\m(\vr,t)$. The index $\m$ denotes the powder ($\p$) or ball ($\b$) and $\vr = (x,y,z)$. The hamiltonian contains some terms representing all relevant interactions working on the matters inside the system as follow, 
\begin{equation}
 H_\m = H_0 + V_{\m-\m} + V_{\m-\v} + V_{\m-\m^\prime} + V_\mathrm{ext} \; ,
\label{eq:h}
\end{equation}
with $\v$ denotes the vial, while $H_0$ is the free matter hamiltonian, that is the kinetic term,
\begin{equation}
 H_0 = \frac{1}{2 m_\m} \sum_{i=1}^{n_\m} \left| \left( \vp_\m \right)_i \right|^2 \; ,
\label{eq:h0}
\end{equation}
where  $n_\m$ is the matter number, $m_\m$ and $\vp_\m$ are the matter mass and momentum. Throughout the paper we assume that the mass or size evolution of matters is uniform for the same matters.

The matter self-interaction $V_{\m-\m}$, the matter--vial interaction $V_{\m-\v}$ and the interaction between different matters may be induced by, for instance,  Coulomb ($V^{\mathrm{Coul}}$) and impact ($V^{\mathrm{imp}}$) potentials,
\begin{eqnarray}
 V^{\mathrm{imp}}_{\m-\mp}(\vr,t) & = & -\sum_{i=1}^{n_\m} \sum_{j=1}^{n_\mp}  \int_0^{\left(\xi_{\m\mp}\right)_{ij}} \d \left(\xi_{\m\mp}\right)_{ij} \, \vn \cdot \left( \vf^{\mathrm{imp}}_{\m\mp} \right)_{ij} \; , 
 \label{eq:vimp}\\
 V^{\mathrm{Coul}}_{\m-\mp}(\vr) & = & Q_\m Q_\mp \, \sum_{i=1}^{n_\m}\sum_{j=1}^{n_\mp} \frac{1}{\left| \left( \vr_\m \right)_i - \left( \vr_\mp \right)_j \right|} \; ,
\label{eq:vcoul}
\end{eqnarray}
with $Q_\m$ is the matter charge, while $\m,\mp : \v, \p, \b$ and $\vn$ is the unit normal vector. These potentials are considered describing the mechanical and static electrical properties of the matters. The first potentials should in fact represent the whole classical dynamics among the matters, i.e. the impact forces among balls and powders. This form will be clarified soon below. The Coulomb potential disappears if the interacting matters have neutral charges. Also it works only in a short range of distance. Therefore, it should be negligible for nanometers scale of physics as in our case. The  impact forces between the vial surface and balls or powders are  treated in the same way using Eq. (\ref{eq:vimp}) by considering that the surface is constructed from a set of fixed spheres \cite{concas}. This is inline with the simulation in the last section where we discretize the vial volume in small spheres with a comparable size as the desired powder size, i.e. few tens nanometers at the largest.

On the other hand, instead of Eq. (\ref{eq:vimp}) we can consider a simpler 'effective potential' like  the  harmonic oscillator $V^{\mathrm{osc}}_{\m-\mp}(\vr) = \frac{1}{2} \, k_{\m\mp} \, \Delta \vr^2$ to represent the whole mechanical properties in terms of 'effective coupling' $k_{\m\mp}$. In this case, if $m_\m \gg m_\mp$, the potential can be rewritten as $V_{\m-\mp} = \frac{1}{2} \, m_\m \, {\omega_\m}^2 \, \Delta \vr^2$ since $\omega_\m \equiv \sqrt{{k_{\m\mp}}/{m_\m}}$. Actually this is the case of ball--powder interaction since $m_\b \gg m_\p$ by the order of namely $O(10^2)$. Nevertheless, in contrast  with its simplicity, $V^{\mathrm{osc}}$ absorbs the time dependency and also interesting physical parameters characterizing the material properties like viscoelasticity, Young modulus etc. The time dependency is important to directly relate the system temperature with physical observables through finite temperature partition function as shown in the next subsection. Therefore, in this paper we take the impact potential to represent the mechanical properties in the system.

The impact potential in Eq. (\ref{eq:vimp}) is induced by the impact force consists of normal and tangential components \cite{concas}, $\vf_{\m\mp}^{\mathrm{imp}}(\vr,t) = \vf_{\m\mp}^N(\vr,t) + \vf_{\m\mp}^T(\vr,t)$. The normal component may be written \cite{brilliantov},
\begin{equation}
 \vf_{\m\mp}^N(\vr,t) = \left [ \frac{2 \Upsilon_{\m\mp}}{3 (1 - v_{\m\mp}^2)} \sqrt{R_{\m\mp}^\mathrm{eff}} \left ( \xi_{\m\mp}^{{3}/{2}} + \frac{3}{2} A_{\m\mp} \sqrt{\xi_{\m\mp}} \, \frac{\d \xi_{\m\mp}}{\d t}\right) \right] \vn  \; .
\label{eq:fn}
\end{equation}
Here the first term is the elastic part based on the Hertz contact law, and the second term is the dissipative one that takes into account material viscosity. $\Upsilon_{\m\mp}$ is the Young modulus and $v_{\m\mp}$ represents the Poisson ratio of the sphere material. The term $R_{\m\mp}^\mathrm{eff} = {(R_\m R_\mp)}/{(R_\m + R_\mp)}$ represents the effective radius, while $\xi_{\m\mp} = R_\m + R_\mp - | \vr_\m - \vr_\mp|$ is the displacement with $R_\m$ is the radius of interacting matter. $A$ is a dissipative parameter \cite{brilliantov,landau,hertzsch}, 
\begin{equation}
A_{\m\mp} = \frac{1}{3} \frac{{3 \eta_\mp - \eta_\m}^2}{3 \eta_\mp + 2 \eta_\m} 
	\left[ \frac{(1 - v_{\m\mp}^2 )( 1 - 2 v_{\m\mp})}{ \Upsilon_{\m\mp} \, v_{\m\mp}^2} \right] \; .
\label{eq:a}
\end {equation}
The viscous constants $\eta_\m$ and $\eta_\mp$ relate the dissipative stress tensor to the deformation tensor \cite{brilliantov,landau}.

There are several proposed formulations for the tangential component $\vf_{\m\mp}^T (\vr,t)$. However it always assumes that the material slide upon each other in the case where the Coulomb condition $\mu \left| \vf_{\m\mp}^N \right| \leq \left| \vf_{\m\mp}^T \right|$ holds, otherwise some viscous frictions occur \cite{saluena}. In particular it follows that $\vf_{\m\mp}^T(\vr,t) \propto m_{\m\mp}^\mathrm{eff}$, where the effective mass is $m_{\m\mp}^\mathrm{eff} \equiv {m_\m m_\mp}/{(m_\m + m_\mp)}$ \cite{concas}. Obviously, in our case with large mass discrepancy between powder and ball, the tangential impact force may be neglected for a good approximation. So, let us from now consider the normal component dominated impact force, that is $\vf_{\m\mp}^{\mathrm{imp}}(\vr,t) \sim \vf_{\m\mp}^N(\vr,t)$. This result simply yields the impact potential as written in Eq. (\ref{eq:vimp}) due to the Euler-Lagrange equation,
\begin{equation}
 \vf = -\frac{\d V}{\d \vr} + \frac{\d}{\d t} \left( \frac{\d V}{\d \vv} \right) \; ,
\label{eq:ele}
\end{equation}
since the dependency on matter velocity appears only in the tangential component $\vf_{\m\mp}^T(\vr,t)$ \cite{concas}.

Beside the interactions among the matters itself, it is also possible to take into account the external potentials working on the whole system. For instance, in the dynamics of ball, the gravitational potential,
\begin{equation}
 V^{\mathrm{grav}}_\mathrm{ext} = m_\m \, G \, \sum_{i=1}^{n_\m} \left( z_\m \right)_i \; ,
\end{equation}
might be important, with $G$ is the gravitational constant. On the other hand, this should be less important for the powder dynamics due to its tiny size. However, we may take into account the effect of external electromagnetic field surrounding the system to the charged matter. The potential is induced by the Lorentz force, $\vf_\m^{\mathrm{EM}} = Q_\m \, (\ve + \vv_\m \times \vb)$, which leads to,
\begin{equation}
 V_{\mathrm{ext}}^{\mathrm{EM}} = Q_\m \sum_{i=1}^{n_\m} \left[ \phi - \left( \vv_\m \right)_i \cdot \va \right] \; ,
\label{eq:vem}
\end{equation}
and satisfies Eq. (\ref{eq:ele}). $\phi$ and $\va$ are the electromagnetic scalar and vector potentials related to the electric and magnetic fields by $\ve = - \vnabla \phi - {\partial \va}/{\partial t}$ and $\vb = \vnabla \times \va$.
The inclusion of electromagnetic potential shifts the kinetic term in Eq. (\ref{eq:h0}) as follow,
\begin{equation}
 H_0 \longrightarrow H_{0 + \mathrm{EM}} = \frac{1}{2 m_\m} \sum_{i=1}^{n_\m} \left| \left( \vp_\m \right)_i - Q_\m \, \va \right|^2 + n_\m \, Q_\m \, \phi \; ,
\label{eq:h0em}
\end{equation}

From now, let us focus only on the dynamics of powder which is our main interest in the sense of comminution process. From Eqs. (\ref{eq:h}), (\ref{eq:h0}), (\ref{eq:vimp}), (\ref{eq:vcoul}) and (\ref{eq:vem}), the total hamiltonian for the powder in our model is,
\begin{eqnarray}
 H_\p & = & \frac{1}{2 m_\p} \sum_{i=1}^{n_\p} \left| \left( \vp_\p \right)_i - Q_\p \, \va \right|^2 + n_\p \,Q_\p \, \phi  
	- \frac{1}{2} \sum_{i\neq j=1}^{n_\p}  \int_0^{\left(\xi_{\p\p}\right)_{ij}} \d \left(\xi_{\p\p}\right)_{ij} \, \vn \cdot \left( \vf^{\mathrm{imp}}_{\p\p} \right)_{ij} 
	\nonumber \\
	&& - \sum_{m:\b,\v}\sum_{i=1}^{n_\p} \sum_{j=1}^{n_\m}  \int_0^{\left(\xi_{\p\m}\right)_{ij}} \d \left(\xi_{\p\m}\right)_{ij} \, \vn \cdot \left( \vf^{\mathrm{imp}}_{\p\m} \right)_{ij} \; ,
\label{eq:hp}
\end{eqnarray}
for $Q_\p \neq 0$. The last two potentials represent the total impact potential among powders; powders and vial; powders and balls respectively. Obviously we do not need to take into account the ball self-interaction $V^{\mathrm{imp}}_{\b-\b}$ nor ball-vial interaction $V^{\mathrm{imp}}_{\b-\v}$. This is actually the advantage of using hamiltonian method.

\subsection{Physical observables}

As mentioned briefly in introduction, the greatest advantage of deploying the hamiltonian method is one can extract some physical observables without solving the EOM's governing the system. Instead, in a canocical ensemble we  consider the partition function of the model,
\begin{equation}
 Z_\m = \int \prod_{i=1}^{n_\m} \d \vp_i \, \d \vr_i \; \mathrm{exp} \left[ 
 -\int_0^\beta \d t \, H_\m
 \right] \; ,
\label{eq:z}
\end{equation}
where $\beta \equiv 1/{(k_B T)}$ with $k_B$ and $T$ are the Boltzman constant and absolute temperature. Having partition function at hand, we can obtain some thermodynamics quantities  in the system through relations,
\begin{equation}
 F_\m = -\frac{1}{\beta} \; \ln Z_\m \; ,
\label{eq:f}
\end{equation}
for free energy and,
\begin{equation}
 P_\m = -\frac{\partial F_\m}{\partial V} = -\frac{F_\m}{V} \; ,
\label{eq:p}
\end{equation}
for pressure in a vial with volume $V$.

In order to see the contributions from the interactions, it is more convenient to consider the normalized partition function,
\begin{equation}
 \bar{Z}_\m \equiv \frac{Z_\m}{Z_{0_\m}} = \frac{\displaystyle \int \prod_{i=1}^{n_\m} \d \vp_i \, \d \vr_i \; \mathrm{exp} \left[ 
 -\int_0^\beta \d t \, H_\m
 \right]}{\displaystyle \int \prod_{i=1}^{n_\m} \d \vp_i \; \mathrm{exp} \left[
 -\int_0^\beta \d t \, H_{0_\m}
 \right]} \; ,
\label{eq:nz}
\end{equation}
and further,
\begin{equation}
 \bar{P}_\m = \bar{F}_\m \equiv \frac{F_\m}{F_{0_\m}} = \frac{\ln Z_\m}{\ln {Z_0}_\m} \; .
\label{eq:nf}
\end{equation}

Performing the integral over time ($t$), we immediately obtain a temperature dependent partition function, while the integrals over $\vp_i$ are the decoupled gaussian integral which can be easily calculated. In the case of Eq. (\ref{eq:hp}) it gives,
\begin{equation}
  Z_\p = \left( \frac{2 \, m_\p \pi}{\beta} \right)^{{n_\p}/2} 
  \int \prod_{i=1}^{n_\p} \d \vr_i  \; \mathrm{exp} \left[ 
 -\int_0^\beta \d t \, H^\prime_\p
 \right] \; ,
\label{eq:z2}
\end{equation}
and, 
\begin{equation}
 \bar{Z}_\p = \int \prod_{i=1}^{n_\p} \d \vr_i  \; \mathrm{exp} \left[ 
 -\int_0^\beta \d t \, H^\prime_\p
 \right] \; ,
\label{eq:zp}
\end{equation}
where the interaction hamiltonian is,
\begin{eqnarray}
 H_\p^\prime & = & n_\p \, Q_\p \, \phi  
	- \frac{1}{2} \sum_{i\neq j=1}^{n_\p}  \int_0^{\left(\xi_{\p\p}\right)_{ij}} \d \left(\xi_{\p\p}\right)_{ij} \, \vn \cdot \left( \vf^{\mathrm{imp}}_{\p\p} \right)_{ij} 
	- \sum_{\m:\b,\v} \sum_{i=1}^{n_\p} \sum_{j=1}^{n_\m}  \int_0^{\left(\xi_{\p\m}\right)_{ij}} \d \left(\xi_{\p\m}\right)_{ij} \, \vn \cdot \left( \vf^{\mathrm{imp}}_{\p\m} \right)_{ij} 
	 \; .
\label{eq:hpp}
\end{eqnarray}
Obviously, only the scalar potential of external electromagnetic field contributes to the  total energy of system under consideration. In other words, we can conclude here that in our model the magnetic field $\vb$ does not influence the ball mill system, but the electric field $\ve$ does.

Moreover, we can perform the integration over time ($t$) and $\xi_{ij}$ to obtain further,
\begin{eqnarray}
 \bar{Z}_\p & = & \int \prod_{i=1}^{n_\p} \d \vr_i  \; \mathrm{exp} \left[ -\beta \left( 
 n_\p \, Q_\p \, \phi  
 - \frac{2}{15} \sum_{i\neq j=1}^{n_\p} \frac{\Upsilon_{\p\p}}{1 - v_{\p\p}^2} \sqrt{R_{\p\p}^\mathrm{eff}} \, \left(\xi_{\p\p}\right)_{ij}^{{5}/{2}} 
  \right.\right. \nonumber \\
  && \left.\left. - \frac{4}{15} \sum_{\m:\b,\v} \sum_{i=1}^{n_\p} \sum_{j=1}^{n_\m} \frac{\Upsilon_{\p\m}}{1 - v_{\p\m}^2} \sqrt{R_{\p\m}^\mathrm{eff}} \, \left(\xi_{\p\m}\right)_{ij}^{{5}/{2}} 
  \right) \right] \; .
\label{eq:zpf}
\end{eqnarray}
From this result, the thermodynamics observables are clearly not affected with the dissipative term, i.e. the second term in Eq. (\ref{eq:fn}). After performing same integration we obtain,  \begin{equation}
  \bar{P}_\p = 1 - \beta \, \cf \,  \ln^{-1} \left( \frac{2 \, m_\p \pi}{\beta} \right) 
  \; ,
\label{eq:pf}
\end{equation}
respectively with,
\begin{eqnarray}
 \cf & \equiv & 2 \,  \int \prod_{i=1}^{n_\p} \d \vr_i  \left[ 
  Q_\p \, \phi  
 - \frac{2}{15 \, n_\p} \sum_{i\neq j=1}^{n_\p} \frac{\Upsilon_{\p\p}}{1 - v_{\p\p}^2} \sqrt{R_{\p\p}^\mathrm{eff}} \, \left(\xi_{\p\p}\right)_{ij}^{{5}/{2}} 
   - \frac{4}{15 \, n_\p} \sum_{\m:\b,\v} \sum_{i=1}^{n_\p} \sum_{j=1}^{n_\m} \frac{\Upsilon_{\p\m}}{1 - v_{\p\m}^2} \sqrt{R_{\p\m}^\mathrm{eff}} \, \left(\xi_{\p\m}\right)_{ij}^{{5}/{2}} 
  \right] \; .
\label{eq:faux}
\end{eqnarray}
Eq. (\ref{eq:pf}) provides a general behavior for temperature-dependent pressure in the model, while the geometrical structure and motion of vial is absorbed in the function $\cf$. From Eq. (\ref{eq:pf}) clearly the physically meaningful regions are for $0 < T < (2 \, m_\p \, \pi \, k_B)^{-1}$ and $T \geq T_\mathrm{th}$. The later is equivalent to the condition,
\begin{equation}
 \cf \leq k_B \, T_\mathrm{th} \, \ln \left( 2 \, m_\p \, \pi \, k_B \, T_\mathrm{th} \right) \, ,
\end{equation}
and $T_\mathrm{th}$ is always greater than $(2 \, m_\p \, \pi \, k_B)^{-1}$.

\section{Summary}

The simulation can in principle be performed to deal with any types of ball mills.  This can be accomplished by replacing the coordinate system of ball mill under consideration represented by $\cf$ in Eq. (\ref{eq:faux}) to the appropriate one which represents its  geometrical motion.

We have proposed and discussed a novel model and approach for comminution processes of nanomaterial using ball mill equipments. The study is focused on investigating the hamiltonian for a ball mill system, and relating it to relevant physical observables through partition function.

Detail simulation and its numerical results are still under progress and will be reported elsewhere.

\begin{theacknowledgments}
The work is supported by the Riset Kompetitif LIPI in fiscal year $2009$. Muhandis thanks the Group for Theoretical and Computational Physics LIPI for warm hospitality during the work.
\end{theacknowledgments}

\bibliographystyle{aipproc}
\bibliography{ballmill}

\begin{thebibliography}{16}
\expandafter\ifx\csname natexlab\endcsname\relax\def\natexlab#1{#1}\fi
\providecommand{\enquote}[1]{``#1''}
\expandafter\ifx\csname url\endcsname\relax
  \def\url#1{\texttt{#1}}\fi
\expandafter\ifx\csname urlprefix\endcsname\relax\def\urlprefix{URL }\fi
\providecommand{\eprint}[2][]{\url{#2}}

\bibitem[Mishra and Rajamani(1992)]{mishra92}
B.~K. Mishra, and R.~K. Rajamani, \emph{Applied Mathematical Modeling}
  \textbf{16}, 598--604 (1992).

\bibitem[Mishra and Rajamani(1994)]{mishra94}
B.~K. Mishra, and R.~K. Rajamani, \emph{International Journal of Mineral
  Processing} \textbf{40}, 171--186 (1994).

\bibitem[Mishra(1995)]{mishra95}
B.~K. Mishra, \emph{Kona Powder Particle} \textbf{13}, 151--158 (1995).

\bibitem[Mishra and Murty(2001)]{mishra01}
B.~K. Mishra, and C.~V.~R. Murty, \emph{Powder Technology} \textbf{115},
  290--297 (2001).

\bibitem[P$\ddot{\mathrm{o}}$schel and
  Salue$\tilde{\mathrm{n}}$a(2001)]{poschel}
T.~P$\ddot{\mathrm{o}}$schel, and C.~Salue$\tilde{\mathrm{n}}$a, \emph{Physical
  Review} \textbf{E64}, 011308 (2001).

\bibitem[Manai et~al.(2002)]{manai}
G.~Manai, F.~Delogu, and M.~Rustici, \emph{Chaos} \textbf{12}, 601--609 (2002).

\bibitem[Davis et~al.(1988)]{davis}
R.~M. Davis, B.~McDermott, and C.~C. Koch, \emph{Metallurgical Transactions}
  \textbf{A19}, 2867 (1988).

\bibitem[Maurice and Courtney(1990)]{maurice1990}
D.~Maurice, and T.~H. Courtney, \emph{Metallurgical Transactions} \textbf{A21},
  289--302 (1990).

\bibitem[Maurice and Courtney(1996)]{maurice1996}
D.~Maurice, and T.~H. Courtney, \emph{Metallurgical Transactions} \textbf{A27},
  1981 (1996).

\bibitem[Delogu et~al.(2000)]{delogu}
F.~Delogu, M.~Monagheddu, G.~Mulas, L.~Schiffini, and G.~Cocco,
  \emph{Innternational Journal of Non-Equilibrium Processing} \textbf{11},
  235--269 (2000).

\bibitem[Wang(2000)]{wang}
W.~Wang, \emph{Modeling and simulation of the dynamics process in high energy
  ball milling of metal powders}, Ph.D. thesis, University of Waikato (2000).

\bibitem[Concas et~al.(2006)]{concas}
A.~Concas, N.~Lai, M.~Pisu, and G.~Cao, \emph{Chemical Engineering Science}
  \textbf{61}, 3746--3760 (2006).

\bibitem[Brilliantov et~al.(1996)]{brilliantov}
N.~V. Brilliantov, F.~Spahn, J.~{M}artin Hertzsch, and
  T.~P$\ddot{\mathrm{o}}$schel, \emph{Physical Review} \textbf{E53}, 5382--5392
  (1996).

\bibitem[Landau and Lifschitz(1965)]{landau}
L.~D. Landau, and E.~M. Lifschitz, \emph{Theory of Elasticity}, Oxford
  University Press, 1965.

\bibitem[Hertzsch et~al.(1995)]{hertzsch}
H.~Hertzsch, F.~sepahan, and N.~V. Brilliantov, \emph{Journal de Physique}
  \textbf{5}, 1725--1738 (1995).

\bibitem[Saluena et~al.(1999)]{saluena}
C.~Saluena, T.~P$\ddot{\mathrm{o}}$schel, and S.~E. Esipov, \emph{Physical
  Review} \textbf{E59}, 4422--4427 (1999).

\end{thebibliography}

\end{document}